\documentstyle[prl,aps,psfig,twocolumn]{revtex}
\begin{document}
\draft
\twocolumn[\hsize\textwidth\columnwidth\hsize\csname
@twocolumnfalse\endcsname
\title{Statistical Features of Drainage Basins in Mars Channel Networks.}
\author{Guido Caldarelli$^1$, Paolo De Los Rios$^2$, Marco Montuori$^1$}
\address{$^1$Sezione INFM and Dip. di Fisica, Universit\`a "La Sapienza", P.le A. Moro 2, 00185 Roma, Italy.}
\address{$^2$Institut de Physique Th\'eorique, Universit\'e de Lausanne, 1015 Lausanne, Switzerland.}
\date{\today}
\maketitle

\begin{abstract}
Erosion by flowing water is one of the major forces shaping the surface of  Earth. Studies in the last
decade have shown, in particular, that the drainage region of rivers, where water is collected, exhibits
scale invariant features characterized by exponents that are the same for rivers around the world. Here we
show that from the data obtained by the MOLA altimeter of the Mars Global Surveyor one can perform the same
analysis for mountain sides on Mars. We then show that in some regions fluid erosion might have played a
role in the present martian landscape.
\end{abstract}
]
\narrowtext

Whether water flowed on the surface of Mars in the past is one of the most intriguing and challenging problems
of contemporary planetary science. Indeed the presence of surface water, even some billion years ago, would make
the odds for the development of life (as we know it) on Mars much higher. At present, the atmospheric conditions
are almost incompatible with surface liquid water, so that its presence in the past has to be detected indirectly,
looking at clues in the geomorphology of the planet.
Moreover, understanding where
water could have been present in the past is of great importance to choose the landing sites of future automatic
and manned missions to Mars.
Already early pictures taken from orbit showed canyons and channels similar to the ones carved by rivers on Earth.
More recently, the high resolution pictures taken in the last three years by the Mars Orbiter Camera (MOC) camera of the
NASA probe Mars Global Surveyor (MGS) have provided a wealth of data that add some more clues: gullies on
the walls of craters and valleys~\cite{ME00_1}, that suggest the presence of liquid water in (geologically) recent times,
sedimentary rock formations typical, on Earth, of the beds of ancient lakes~\cite{ME00_2}, the presence of an ocean
in the northern hemisphere~\cite{Head99} and rootless cones similar to the ones that are found on Earth where molten lava
has flowed over waterlogged ground\cite{Lanagan01}.
All these photographic evidence needs to be complemented by some more robust analysis on data such as the ones
from the Mars Orbiter Laser Altimeter (MOLA).
Here we show that the channel networks that can be inferred
from the MOLA data bear similarities with the structures carved by water erosion on Earth.

On Earth fluvial systems can be described through the self-affine properties of their drainage basin, where water,
either from underground sources or from rain, is collected~\cite{RIR,TBR88}. Those self-affine properties have been established
starting from  Hack's seminal work where the scaling relation (\ref{Eq1}) between the basin area $A$ and the
length $L_{\parallel}$ of the
basin's main stream was proposed~\cite{Hack}:
\begin{equation}
A^h \sim L_{\parallel}
\label{Eq1}
\end{equation}
>From measures on real rivers, the Hack's exponent $h$ takes values between
$0.5$ and $0.7$. The allometric relation between $A$ and $L_{\parallel}$ is just one in a number of scaling
properties of drainage basins.
In general, the drainage basin is characterized by the longitudinal length $L_{\parallel}$
(we use the same symbol as for the main
stream length since, for real rivers, they scale in the same way) and by the transverse one,
$L_\perp$ in such a way that $L_\perp \sim L_{\parallel}^H$.
If the exponent $H$ is less than $1$, the basin is self-affine (the width of the basin scales slower than its length).
If the basin is self-similar $H$ is equal to $1$ (the trivial case where the width scales as the length). The
two exponents $H$ and $h$ are connected, as it is easy to see  by comparing $L_\perp \sim L_\parallel^H$
together with  Hack's law (\ref{Eq1}): one finds
that $h=1/(1+H)$.  These two macroscopic features are not the only ones available in the study of drainage basins.
In order to describe even more precisely the statistical properties channel networks, geomorphologists use Digital
Elevation Models (DEM)~\cite{Band,DWMM92}. Measures, either ground based or, in more recent times, by satellites, provide the
average height of an area that, on Earth, can be as little as $30 \times 30 m^2$. Each of these square units is associated
to a pixel on an image (or, technically, to a site on a two dimensional square grid). Then, water collected in
each pixel flows toward the lowest of the neighbor pixels, according to a simple maximal slope rule
(other rules would be also viable, such as, for example, that water flows toward more than a single neighbor,
and is partitioned proportionally to the slopes). If the maximal slope rule is applied, this procedure produces
a branched structure without loops that should reproduce the visible river network. In a loopless branched structure,
it is easy to define the region upstream of a given point, that is, the region whose collected water will all flow to
that point. Therefore it is possible to label each and every point on the map by the area $A$ of its own drainage basin.
The final outlet is of course labeled by the whole area of the basin. It is possible to draw the histogram of these
areas for a given basin, finding that the frequency $P(A)$ to find a sub-basin of area $A$ follows the law $P(A) \sim A^{-\tau}$,
where $\tau = 1.43 - 1.46$. On the same branched structure, it is possible to measure the upstream length $L$ from a given point,
as the distance of that point from the furthest source (a source is defined as a point without any upstream basin,
that is, that does not collect any upstream water). Again, these lengths can be organized in a histogram, whose asymptotic
behavior for the frequency distribution  $\Pi(L)$ behaves as $\Pi(L) \sim L^{-\gamma}$, where
$\gamma = 1.7 - 1.8$. All these exponents, and others
that could be defined, depend on the above exponent $H$ relating $L_\perp$ and $L_\parallel$, and then
depend ultimately on the fractal properties
of the basin. Using finite size scaling arguments, Maritan {\it et al.}~\cite{Maritan96} have shown that
$\tau = 2-h$, and $\gamma = 1/h$. These relations
hold for any kind of branched structure, whatever its origin, and they show that indeed there is only one independent exponent
that determines all the others. Actually, one more independent exponent is the fractal dimension of the main stream; yet,
river networks are, on the average, directed because water flows down slopes, which gives a preferential direction. Consequently the
fractal dimension of the main stream is $d_l=1$. Still, the relations do not tell why the exponents should take a particular
set of values.
One can study the same statistical properties on a random landscape. In this case of course, due to the nature of the
landscape, many outlets (points lower than all of their neighbors) will be present in the system. We will refer hereafter to
these points as "pits". These points act as sinks where water is only collected and not distributed around. This effect
is so strong, that aggregation of rivers does not take place. The drainage basins have a characteristic small size and the
above distributions $P(A)$ and $\Pi(L)$ show an exponential decay. In order to remove inner pits, one can recursively raise
them at the height of the lowest neighbor (therefore simulating the behavior of water in a real lake when the liquid level
increases steadily). By this procedure, one finally deals with connected structures (maybe with multiple outlets on the boundaries)
whose landscape is very different from the original one (typically one can expect an overlap of about $20-25\%$ between the
original and the final landscape). The spanning tree formed by the river has peculiar statistical properties listed in Table I.
This class of spanning trees has been also studied analytically  and will be hereafter described as the class of the
"random spanning trees"~\cite{Coniglio89}. Since real rivers are not described by this class of spanning trees, it has been conjectured
that an optimization process took place in order to shape the drainage basins in their present form. In particular with the
aid of the Optimal Channel Networks (OCN)~\cite{RIturbe92} model one can show in a rigorous
way that by requiring minimization of the
total gravitational energy dissipation in the system one can reshape a random spanning tree, transforming it into one with the
same statistical properties of the real basins. We then want to exploit this difference in order to check if the martian
landscape has ever been sculpted by fluvial erosion.

\begin{figure}
\centerline{\psfig{file=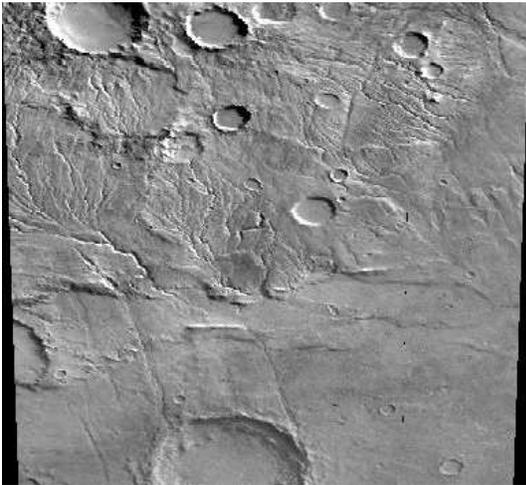,width=7.0cm}}
\caption{Viking Image for the Warrego Vallis (around $41^\circ$S, $91^\circ$W);
some "river-like"structures are present on the center-left
(courtesy from NASA, PDS Planetary Image Atlas).}
\label{Fig1}
\end{figure}
\begin{figure}
\centerline{\psfig{file=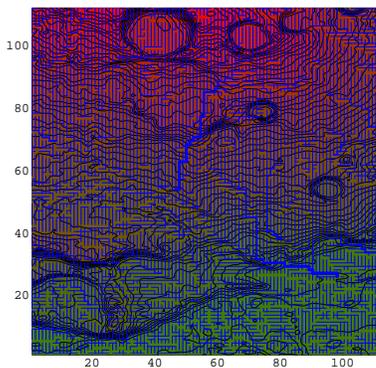,width=5.0cm}}
\caption{Computer reconstruction of the landscape in the area of Warrego Vallis;
the main crater on the top left of Fig.1 is at the center of the red area.}
\label{Fig2}

\end{figure}
\begin{figure}
\centerline{\psfig{file=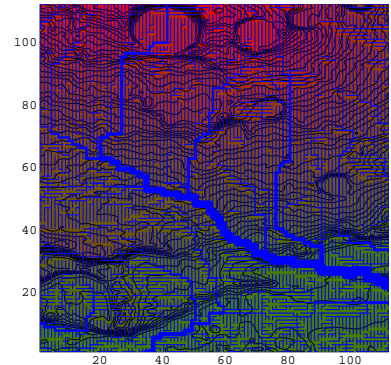,width=5.0cm}}
\caption{Computer reconstruction of the landscape in the area of Warrego Vallis, after
pit removal. The reconstructed river structure is "reasonably" similar to the visible one from
Fig.1. Moreover it is evidently composed of patches of the channels in Fig.2}
\label{Fig2bis}
\end{figure}
\begin{figure}
\centerline{\psfig{file=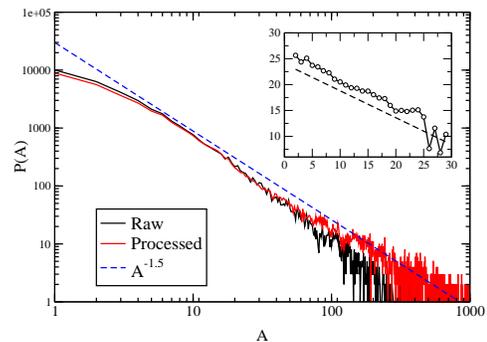,width=7.0cm}}
\caption{Histogram of the areas for the Warrego Vallis. Inset: the log-log plot of
same histogram, binned over area intervals of ratio $1.5$; the dashed line has slope $0.5$.}
\label{Fig3}
\end{figure}

We then performed the same analysis on the surface of Mars. The latest MOLA
data provide us with a DEM such that every pixel covers a surface of about
$1500\times 1500 m^2$ (roughly $0.03125^{\circ} \times 0.03125^{\circ}$). With these data at hand, we have
chosen four regions of about $100\times 100 km^2$, where to reconstruct a network from a DEM:
the {\em Warrego Vallis} shown in Fig.\ref{Fig1}, where some river-like structures are visible
in both Viking Orbiter and MOC pictures; {\em Solis Planum} and {\em Noctis Labyrinthum},
two rough districts in the Tharsis region; and a region from the northern hemisphere,
where an ocean could have existed in the past~\cite{Head99}. On these regions we reconstructed
a channel network using the maximal slope rule: the results are shown in Fig.\ref{Fig2} where
we represented every pixel by the area it collects (at this level we still do not
speculate whether such an area corresponds to any collected water). In Fig.\ref{Fig3} we show
the statistical analysis of the areas: interestingly, the histogram of the areas for the
flat northern lowlands show a clear exponential decrease, corresponding to the absence
 of any particular correlation in the network, and in agreement with random surfaces; on
 the contrary, the histograms for the other three regions clearly exhibit a power-law decay,
 consistent with the existence of correlated structures in the channel networks. The exponents $\tau$
 that can be fitted are between $1.7$ and $1.9$, with a quite large error-bar due the poverty of the
 statistics, yet unambiguously larger than $1.6$ and smaller than $2$. The upstream length histogram
 has also been collected, giving exponents $\gamma$ around $1.7$.
Some observations are already possible: the exponential distribution of the northern hemisphere
areas is indicative of the absence of long-range correlations in the system, as we would expect
on a rough, random surface (and indeed this is the result on an artificially generated random landscapes);
 on the other hand, the power-law distributions in the other regions are suggestive that some correlation
 building process has taken place. Still, the exponents are larger than those of river networks on Earth,
 and moreover, the relation $\tau=2-1/\gamma$ is not respected. Actually, the above scaling relations have been
 obtained in the single outlet approximation. If many inner outlets are present, no definite
relation has been found.
As mentioned above, we can try to eliminate pits from the landscape, letting water fill them up to the
level of the lowest neighbor, and then flowing toward it. Repeating this procedure for all the inner pits,
 the only possible outlets are located on the boundaries. This kind of reconstruction is of course crude,
  and it implies an extrapolation about some past behavior. Other reconstructions could be performed,
  since we are trying to recover some features that formed billions of years ago, and that have been
  surely smoothed by atmospheric erosion. Yet, we feel that, in the hypothesis that some form of fluid
  (be it water or liquid $CO_2$) ran along martian channels, the pit-filling procedure is reasonable, and
  subject to simple laws. Moreover, a simple criterion can be used to decide how much this procedure
  changes the network: we define the overlap between the processed and unprocessed networks as the
  fraction of sites that do not change their outflow direction. A high value of the overlap means
  that filling pits does not change much the network, whereas a low overlap implies that the network
  has been completely modified.  Using this criterion, we can see that a random network changes almost
  completely upon processing, whereas the three channel networks have large overlaps, implying that no
  great rearrangements have taken place, and that therefore it is tempting to associate such networks
  to real rivers. In Figs.\ref{Fig2} and \ref{Fig2bis}, the raw network and the reconstructed one are shown: first,
  it is clear enough that large pieces of the raw network are still present after the pit-filling
  procedure has been applied; moreover, there is a good visual correlation between the reconstructed
  network and the one from orbit pictures. The exponents are also shown in Table I, and they are
  clearly approaching the exponents of Earth river networks. Such similarity suggests that a
  common mechanism could have produced both: we know that on Earth such structures are formed by
  water erosion.

In comparison with data available on Earth ($30\times 30 m^2$), the resolution available at present is extremely
low. Yet, we obtained the same results working on maps of $0.0625^{\circ} \times 0.0625^{\circ}$, the highest resolution
available before March 2001. Moreover, as the Mars Global Surveyor covers more and more orbits, the
MOLA also collects more and more data so that a reliable average (for which at least four measures
are needed) can be obtained for areas increasingly smaller: at the present rate, by the time the
instrument will be shut off the resolution could of the order of $100 \times 100 m^2$  or less (for a real
time count of the laser shots, look at {\em http://sebago.mit.edu/shots//}), so that detailed digital
elevation maps comparable with the one for Earth could be processed. A problem related to the
resolution is also the appropriate way to redistribute the collected areas to downstream sites.
All through this work we used the maximal slope rule; yet, as mentioned above, other rules could
also be used. In particular, we tried to partition areas proportionally to the slope. Such a rule
is reasonable when the resolution is still not high enough, so that water could flow along more
than a single direction. In all the examined cases we always found an exponential behavior of the
distributions $P(A)$ and $\Pi(L)$. Anyway such a rule should be less and less appropriate as the resolution
increases, so that, as better data will be available, the maximal slope rule should become the correct
one to be used. A further problem with the interpretation of the MOLA data is that we are trying to
look at channel networks produced maybe billions of years ago. Over such a huge period,  winds in
the thin martian atmosphere could have smoothed the landscape~\cite{PMD01}. Going back to the original
landscape is extremely difficult; erosion rates can be greatly varied by climate changes.
Nevertheless, a simplistic idea could be to apply some anti-diffusion, but the starting
landscape has a resolution such that we could not be sure about the outcome (anti-diffusion is an
unstable process). Yet, we can try to look at the behavior of an even more eroded landscape, to
understand how much erosion influences the area and upstream length statistics. Applying some
diffusion (that, on the contrary, is a stable process), we find smoother and smoother surfaces,
yet the statistics do not change considerably. If we extrapolate back in the past the same
considerations, we can conclude that today we are looking at statistics close to the ones when
the networks have been produced.

The results of this work are of course speculative, yet, since the debate on whether water
ever flowed on Mars is extremely hot, we strongly believe that as many clues as possible
should be examined, waiting for the time when some more specific mission (either robotic or,
in a more distant future, manned) will give some definitive answers on the questions that
we can ask on such indirect and general observations that we have at hand right now. Answering
the question on whether water ever flowed on Mars is of extreme importance: Mars is the closest
place that could have harbored life in the past, close enough that, at some point in the near
future, we could have direct evidence of it. Moreover, understanding where water was present in
the past could also be of relevance in choosing the landing site for future human missions:
the site should be as much scientifically interesting as possible and, maybe, rich in natural
resources that could be used by astronauts. The present study tries indeed to point out the
 possible scientific merit of some regions, that could also be associated with  some resource
 richness, as river networks usually are on Earth.

We thank the MOLA Team at NASA for making the MOLA data readily available online,
and for helping us to extract the information for our analysis.
This work has been partially supported by the EU Network ERBFMRXCT980183.

\begin{table}[ht]
\begin{center}
\begin{tabular}{|l|c|c|c|}
Surface & H & $\tau$ & $\gamma$ \\
\hline
Earth & $0.7-0.8$ &  $1.43-1.46$ & $1.7-1.8$  \\
\hline
Mars & $\sim 0.7$ &  $ \sim 1.8$ &  $ \sim 1.7$ \\
\hline
Mars ($85-90$\%) & $\sim 0.6$ & $\sim 1.5$ & $ \sim 2.0$ \\
\hline
Mars North. Hemisphere & n.d. &  Exp. &  Exp. \\
\hline
Random Surf. & n.d. &  Exp. &  Exp. \\
\hline
Random Surf. ($20-25$\%) & $1$ & $11/8$ & $5/4$ \\
\end{tabular}
\caption{Exponents for Earth river networks, for Mars channel networks
(processed and unprocessed), for the martian northern hemisphere
(Exp.=exponential; n.d.=not defined), and for a random surface
(processed and unprocessed); in brackets the overlap is reported.}
\label{Tab1}
\end{center}
\end{table}


\end{document}